\documentclass[a4paper,fleqn,usenatbib]{mnras}

\usepackage{newtxtext,newtxmath}
\usepackage[T1]{fontenc}
\usepackage{ae,aecompl}

\usepackage{graphicx}	% Including figure files
\usepackage{amsmath}	% Advanced maths commands
\usepackage{amssymb}	% Extra maths symbols

\title[A1689 in Modified Gravity, max. 45 characters]{Galaxy Cluster A1689 in Modified MOND, MOG and Emergent Gravity}

\author[A. O. Hodson et al]{
Alistair O. Hodson,$^{1}$\thanks{E-mail: aoh2@st-andrews.ac.uk (AOH)}
Hongsheng Zhao$^{1}$
\\
$^{1}$School of Physics and Astronomy, University of St Andrews, Scotland\\
}

\date{Accepted XXX. Received YYY; in original form ZZZ}

\pubyear{2015}

\begin{document}
\label{firstpage}
\pagerange{\pageref{firstpage}--\pageref{lastpage}}
\maketitle

% Abstract of the paper
\begin{abstract}
We model the cluster A1689 in two modified MOND frameworks (EMOND and what we call generalised MOND or GMOND) with the aim of determining whether it is possible to explain the inferred acceleration profile, from gravitational lensing, without the aid of dark matter. We also compare our result to predictions from MOG/STVG and Emergent Gravity. By using a baryonic mass model, we determine the total gravitational acceleration predicted by the modified gravitational equations and compare the result to NFW profiles of dark matter studies on A1689 from the literature. Theory parameters are inferred empirically, with the aid of previous work. We are able to reproduce the desired acceleration profile of A1689 for GMOND, EMOND and MOG, but not Emergent Gravity. There is much more work which needs to be conducted with regards to understanding how the GMOND parameters behave in different environments. Furthermore, we show that the exact baryonic profile becomes very important when undertaking modified gravity modelling rather than $\Lambda$CDM.
\end{abstract}

% Select between one and six entries from the list of approved keywords.
% Don't make up new ones.

%%%%%%%%%%%%%%%%%%%%%%%%%%%%%%%%%%%%%%%%%%%%%%%%%%

%%%%%%%%%%%%%%%%% BODY OF PAPER %%%%%%%%%%%%%%%%%%

\section{Introduction}

The standard paradigm for explaining the mass budget of galaxies and galaxy clusters is cold dark matter. Despite the successes of this paradigm, it is interesting to research possible alternatives focussed on modified gravity, determining whether it is possible explain galaxies and galaxy clusters without this elusive dark matter. This can be a challenging endeavour. Particularly, explaining galaxy cluster mass profiles without dark matter of any kind is a difficult task. One of the most popular gravitational paradigms which does not assume dark matter a priori is Modified Newtonian Dynamics (MOND) \citep{milgrom19831,milgrom19832,milgrom19833,bekenstein1984} which modifies the Newtonian gravitational law in low acceleration environments, determined by an acceleration scale $a_{0}$. The MOND paradigm, which has success on the galaxy scale, is not compatible with galaxy clusters. Neutrinos have been proposed as a solution to this problem, with mixed results. For a historical perspective on clusters with neutrinos in the context of MOND, the authors refer the reader to works such as \citep{sanders1999,sanders2003,angus20081,angus2009,angusneutrino1,angusneutrino2}. For a more general review of MOND, the authors refer the reader to \cite{famaeyreview}.

The reason why MOND does not explain galaxy clusters well is that the gravitational acceleration within these objects are high and thus the MOND effect on the governing gravitational equations is weak. Perhaps the solution to reconcile MOND with galaxy clusters, without dark matter, is to generalise the MOND formulation. One attempt to do this  was Extended MOND (EMOND) \citep{EMOND}, which made the MOND acceleration scale, $a_{0}$ a function of gravitational potential, $A_{0}(\Phi)$, such that the acceleration scale is larger in clusters. Enhancing the acceleration scale allows deviations from Newtonian gravity to occur in higher acceleration environments (e.g. clusters) with the aim to rectify the issues MOND has in galaxy clusters. The EMOND paradigm has been tested with a sample of galaxy clusters in \cite{HodsonEMOND} and in the context of ultra-diffuse galaxies in Hodson and Zhao 2017 (submitted) (Hereafter HZ2017).

Another, yet untested, version of MOND which aims to explain galaxy clusters is the work of \cite{SF1} which, for the purposes of this work, we will call generalised MOND or GMOND. The underlying idea of GMOND is the modified gravitational law is governed by two, non-constant parameters which define the transitions between Newtonian physics, MOND-like physics and a third regime, which is compatible with galaxy clusters. The main reason why this paradigm has, as of yet, not been tested is that there does not exist a dynamical mechanism which controls the values of the two parameters, which govern the gravitational law in different environments\footnote{Suggestions were made in \cite{SF1} as to what the parameters should approximately be in galaxy clusters and galaxies.}. Before a dynamical mechanism can be determined, constraints on the parameters need to be determined from fitting individual astrophysical objects and looking for trends. As the paradigm is a subset of MOND, there are obvious constraints already in place, mainly that galaxies should be in the MOND regime of the equations. The main constraints which need determined are with regards to galaxy clusters.

The MOND paradigm is not the only attempt to reconcile galaxy and extragalactic dynamics without dark matter. Other ideas include Modified Gravity or Scalar-Tensor-Vector-Gravity (MOG/STVG) \citep{STVG}, which makes use of additional scalar and vector fields to boost gravity, and Emergent Gravity (EG) \citep{EG} which makes use of thermodynamics and quantum information theory. The MOG paradigm has undergone less rigorous testing than MOND but has enjoyed success with, for example, rotation curves of galaxies \citep{MOGrotation} and mass profiles of galaxy clusters \citep{MOGclusters}\footnote{Investigations into understanding the lensing signature of the bullet-cluster have also been made \citep{MOGbullet} but have not been published yet.}. The EG framework is very recent and thus had little chance to be tested properly. However, it has been tested in dwarf galaxies in \cite{EGdwarf} and with gravitational lensing prediction in \cite{EGlensing}, enjoying success in both studies. However, recent work discussing rotation curves and solar system constrains \citep{EGRot} as well as the radial acceleration relation \citep{EGRAR} have shown tension between EG and observation. Work into galaxy clusters has been conducted as well \citep{EGcluster}.

Recently, work on galaxy cluster A1689 showed that the currently explored gravitational paradigms such as MOND, EG and MOG cannot explain the lensing inferred mass without the aid of some kind of cluster dark matter, such as neutrinos \citep{A1689Paper}.

It seems fitting to determine whether the EMOND and GMOND paradigm of \cite{SF1} can explain the inferred mass profile of cluster A1689, without resorting to dark matter. We also explicitly show the results for MOG and EG as our model is slightly different to that of \cite{A1689Paper} and thus the results of \cite{A1689MOG}, which looked at A1689 in MOG, will change. 

This paper is organised as follows. In Section \ref{NewtModel} we describe how A1689 has been modelled in Newtonian and show the dark matter profile to which we compare the modified gravity theories. In Section \ref{EMONDSec} we review the MOND and EMOND paradigms. In Section \ref{MONDSec}, we review GMOND i.e. the work of \cite{SF1}.  In Section \ref{MOGSec} we show the MOG spherical force law. In Section \ref{EGSec} we outline the EG paradigm. We briefly summarise our relevant equations in Section \ref{SummarySec}. The baryonic mass model which we adopt is outlined Section \ref{A1689}. We also compare to the model of \cite{A1689Paper} and explain why we choose a different model. Our results are shown in Section \ref{Results}. We finally conclude in Section \ref{Conclusion}. 

\section{Newtonian  Modelling}\label{NewtModel}

The total gravitational acceleration in $\Lambda$CDM for galaxy clusters can be simply summarized as

\begin{equation}
\nabla\Phi = \nabla\Phi_{DM} + \nabla\Phi_{b}
\end{equation}

\noindent where the subscripts DM and b are the dark matter and baryon contributions respectively. The baryons are composed primarily of gas and galaxies. We will discuss the mathematical model for these components in Section \ref{A1689}.

The galaxy cluster A1689 has undergone rigorous modelling in previous works such as \cite{A1689Broad,A1689Hal,A1689UmBroad,A1689Coe,A1689Um}. We therefore have an idea of the theorised amount of dark matter in the system. In the $\Lambda$CDM paradigm, the total amount of gravity is the sum of the baryons and the dark matter. In galaxy clusters, the main baryon contribution to the centre of the cluster is the galaxies and in the outer regions, the intra-cluster gas becomes dominant. The gas contribution is well constrained by X-ray observations, whereas the galaxy contribution is somewhat harder to constrain. We will discuss the parameterisation of the galaxies and the gas in Section \ref{A1689}.  

The main source of gravity, which almost entirely dominates over the baryons is the dark matter. Commonly, dark matter is modelled in galaxy clusters using a Navarro-Frenk-White (NFW) profile \cite[][]{NFWpaper,zhaoprofile}, which is motivated from large scale simulations. The NFW density profile is described via,

\begin{equation}
\rho_{NFW}(r) = \frac{\rho_{s}}{\frac{r}{r_{s}}\left( 1 + \frac{r}{r_{s}} \right)^{2}}
\end{equation}

\noindent or in terms of mass,

\begin{equation}
M_{NFW}(r) = 4 \pi \rho_{s} r_{s}^{3} \left[  \log \left( \frac{r_{s} + r}{r_{s}} \right) -  \frac{r}{r_{s} + r}\right].
\end{equation}

\noindent where $\rho_{s}$ and $r_{s}$ are the scale density and radius of the dark matter halo respectively, which are the free parameters. However, when describing dark matter haloes, it is usual to write the free parameters in terms of the total mass at radius $r_{200}$\footnote{$r_{200}$ is the radius at which the average density (Mass/Volume) reaches 200 times the critical density of the universe.} and the concentration $c_{200}$, which are linked to $\rho_{s}$ and $r_{s}$. The radius, $r_{200}$ is determined by solving

\begin{equation}
\frac{3 M_{200}}{800 \pi r_{200}^{3}} = \rho_{c} 
\end{equation}

\noindent where $\rho_{c} = 3 H^{2}(z)/(8\pi G)$ is the critical density of the universe at the given redshift, $z$ of the cluster. The parameter $r_{s}$ is then calculated via

\begin{equation}
r_{s} = \frac{r_{200}}{c_{200}}.
\end{equation}

\noindent Finally, $\rho_{s}$ is determined by solving the equation

\begin{equation}
M_{200} = 4 \pi \rho_{s} r_{s}^{3} \left[  \log \left( \frac{r_{s} + r_{200}}{r_{s}} \right) -  \frac{r_{200}}{r_{s} + r_{200}}\right].
\end{equation}

We then have a well defined dark matter halo model. The parameters which were derived from the literature are summarised in Table 9 of \cite{A1689Um}, where we focus on the spherical models.

\section{MOND and EMOND dynamics}\label{EMONDSec}

\subsection{MOND}

Before looking at the EMOND equations, we should review the original MOND formulation. The MOND paradigm alters the Newtonian Poisson equation as,

\begin{equation}\label{MONDPoisson}
\nabla \cdot \left[ \nabla \Phi ~\mu\left( \frac{|\nabla \Phi|}{a_{0}} \right) \right] = 4\pi G \rho_{b}
\end{equation}

\noindent where $\Phi$ is the total gravitational potential, $\nabla\Phi$ is the total gravitational acceleration, $\mu(x)$ is the MOND interpolation function where $x \equiv \nabla\Phi/a_{0}$, $a_{0} \approx 1.2 \times 10^{-10}$ ms$^{-2}$ is a constant acceleration scale and $\rho_{b}$ is the baryonic matter density. The interpolation function, along with $a_{0}$, controls the deviation from Newtonian dynamics. The functional form of $\mu(x)$ is chosen such that $\mu(x)\approx 1$ when $x \gg 1$ and $\mu(x)\approx x$ when $x \ll 1$. This means that when when the internal gravitational acceleration is high, Eqn \ref{MONDPoisson} behaves like the Newtonian equivalent. When the acceleration is low, the gravitational acceleration is enhanced compared to the Newtonian prediction.

Equation \ref{MONDPoisson} has enjoyed some success in explaining dynamics on the galaxy scale without including a dark matter component to the matter density. The realisation that Eqn \ref{MONDPoisson} cannot explain galaxy clusters, due to the MOND enhancement being insufficient to explain the internal dynamics, without the inclusion of dark matter, has prompted investigations into alternate and generalisations of the MOND paradigm. In this work, we will discuss two modified MOND frameworks, EMOND and what we call generalised MOND (GMOND). 

\subsection{EMOND}

As mentioned, EMOND is an extension to the MOND paradigm such that the acceleration scale, $a_{0}$, is altered from being constant to being a function of gravitational potential. The EMOND paradigm was first proposed by \cite{EMOND}, tested with galaxy clusters by \cite{HodsonEMOND} and then applied to a sample of ultra-diffuse galaxies by HZ2017. It seems fitting to determine whether the same EMOND formulation used in the previous works can explain the observed properties of A1689. We now briefly outline the equations of EMOND.

The EMOND Poisson equation was determined to be \citep{EMOND},

\begin{equation}\label{EMONDPoiss}
4 \pi G \rho_{b} = \nabla \cdot \left[\mu\left( \frac{|\nabla \Phi|}{A_{0}(\Phi)}  \right) \nabla \Phi  \right] - T_{2},
\end{equation}
where
\begin{equation}
T_{2} = \frac{1}{8\pi G}\left| \frac{d (A_{0}(\Phi))^{2}}{d\Phi} \right|\left[ y F'(y) - F(y) \right]
\end{equation}

\noindent and, in this case,  $dF(y)/dy = \mu(\sqrt{y})$ where $y = |\nabla \Phi|^{2}/A_{0}(\Phi)^{2}$. In this equation, $\rho_{b}$ is the baryonic matter density and $\Phi$ is the total gravitational potential. For galaxy clusters, it was found that the $T_{2}$ term is negligible \citep{HodsonEMOND} and therefore the EMOND Poisson equation can be integrated in the same way as the regular MOND Poisson equation to find a simple, spherical approximation,

\begin{equation}\label{EMONDPoiss2}
\nabla\Phi_{N} = \mu(\sqrt{y})\nabla\Phi
\end{equation}

\noindent where we will adopt a small change  to the simple interpolation function
\citep{simplemu},
 %$\mu(\sqrt{y}) = \sqrt{y}/(1+\sqrt{y})$ 
\begin{equation}\label{simple}
\mu(x) = \max\left[  \frac{x}{1+x},  \frac{\epsilon}{1+\epsilon}\right]
\end{equation}
where $\epsilon$ is a small parameter.

The last aspect of the EMOND modelling is the functional choice for $A_{0}(\Phi)$. For this work, we use a function,

\begin{equation}\label{A0tan}
A_{0}(\Phi) =  \frac{a_{0}}{\epsilon}\mu\left[ \left( \frac{\Phi}{\Phi_{0}} \right)^{2q} \right]
\end{equation}
which means
\begin{equation}\label{simple}
A_{0}(\Phi) = \frac{a_{0}}{\epsilon} \max\left[ \frac{|\Phi|^{2q}}{|\Phi_0|^{2q}+|\Phi|^{2q}}, \frac{\epsilon}{1+\epsilon} \right]
\end{equation}
where $\epsilon$ is the same small parameter. For our cluster, $x_{A0} \equiv \left( \frac{\Phi}{\Phi_{0}} \right)^{2q}$ is always greater than $\epsilon$, so the dynamics are equivalent to the pure simple interpolation function and the analysis in HZ2017. The small parameter is not free, but fixed by the maximum value which we let our acceleration scale to take, $A_{0~ \rm max}$, $\epsilon = a_{0}/A_{0~ \rm max}$.

\noindent This function also gives reasonable results for a sample of galaxy clusters (HZ2017). The maximum value of the EMOND acceleration scale we choose to be $A_{0~max} \approx 100 a_{0}$. Other parameters in EMOND are $\Phi_{0} \approx -3800000^{2}$ m$^{2}$s$^{-2}$, a scale potential analogous to $a_{0}$ in regular MOND, and $q=1$, a dimensionless parameter which controls the sharpness of the transition from high to low values of $A_{0}(\Phi)$\footnote{These parameters are not fixed at the values given, but are consistent with the values used in HZ2017.}.

\section{Introduction to generalised MOND}\label{MONDSec}

\subsection{GMOND equations}

Next we will discuss the modified gravity law of \cite{SF1}, which we call GMOND. 
\cite{SF1} proposed the existence of a scalar field, $\phi$, such that 
\begin{equation}\label{Total_Potential}
\Phi = \Phi_{N}  +  \tilde{\phi}, ~~\tilde{\phi} \equiv - \frac{\phi}{M_{\rm pl}}
\end{equation}
where $\phi$ is a positive quantity with unit $\rm kg~ m^{2}~s^{-2}$, thanks to the reduced Planck mass  $M_{\rm pl} \equiv \sqrt{\frac{\hbar c}{8 \pi G}}$.

The main attraction of GMOND is its proposal that there is a third regime, apart from Newtonian and deep-MOND, to model galaxy clusters, governed by a parameter, $f$. 
So 
\begin{equation}\label{Justin_MOND_Cases}
\nabla\Phi_{N} = \nabla\Phi - \nabla \tilde{\phi} = \begin{cases}

  \frac{\nabla\tilde{\phi}}{A_{0}} \nabla\tilde{\phi} & \text{for $\frac{A_{0}}{f} \ll  \nabla\tilde{\phi}$ (in Galaxies)} \\
  
  \frac{1}{f} \nabla\tilde{\phi} & \text{for $ \nabla\tilde{\phi} \ll\frac{A_{0}}{f}$~ (in Clusters).}

\end{cases}
\end{equation} 
In Equation \ref{Justin_MOND_Cases}, $A_{0}$, with units of acceleration, and the dimensionless parameter, $f$, are ideally dynamical quantities which take different values depending on the type of system being modelled. However, no dynamical mechanism has been proposed for these parameters yet.

A modified Poisson equation for this paradigm was proposed, derived from a Lagrangian, to model the scalar field

\begin{equation}\label{Justin_Poisson}
\nabla \cdot \left[ F'(Y) \nabla \tilde{\phi} \right] = 4 \pi G \rho_{b}, \qquad Y \equiv - \frac{(\nabla \tilde{\phi})^{2}}{A_{0}^{2}}.
\end{equation}
where, e.g., 
\begin{equation}\label{Ffunc}
F(Y) = \frac{Y}{f}\sqrt{1 - \left( \frac{2f}{3} \right)^{2} Y}.
\end{equation}

Equation \ref{Justin_Poisson} is similar to the MOND Poisson equation with a different interpolation function. Note $F'(Y) \rightarrow 1/f$ for small $\sqrt{-Y}\ll 1/f$, and $F'(Y)\rightarrow \sqrt{-Y} \gg 1/f$ for large $Y$.

Equation \ref{Justin_Poisson} is a generalisation of the MOND paradigm which, in theory, can model astrophysical systems of all scales without the need for dark matter. Currently there has been no astrophysical testing or modelling of this paradigm. The main issue with the paradigm at this stage is the lack of a dynamical mechanism for controlling $A_{0}$ and $f$. In our work, we do not attempt to make these parameters dynamical, but rather propose constraints on them if they were made to be dynamical. 

Finally, Eqn \ref{Justin_Poisson} has a fully analytic scalar field gradient, $\nabla \phi$, in spherical symmetry (see Appendix \ref{AppAnalytic}). Therefore, given a spherical baryonic mass distribution, it is possible to make GMOND predictions analytically.

\subsection{Linking EMOND and GMOND}

Although different in formalism, EMOND and GMOND are actually linked. In EMOND, we make the MOND acceleration scale dynamical by introducing a dependence on gravitational potential. In our example here, we define the interpolation function such that it has a limit $\mu(\sqrt{y}) = \epsilon/(1+ \epsilon)$ for low values of $\sqrt{y}$, where $\sqrt{y} \equiv \frac{\nabla\Phi}{A_{0}(\Phi)}$. Therefore the EMOND equation becomes $\nabla\Phi = \nabla\Phi_{N}(1+\epsilon)/\epsilon$. This looks like the cluster regime of GMOND, where $(1 + \epsilon)/\epsilon = f$. In GMOND, $\epsilon$ would be dynamical as well, whereas is it fixed in  EMOND.

\section{Modified Gravity (MOG)}\label{MOGSec}

Like MOND, MOG, sometimes known as Scalar-Tensor-Vector Gravity (STVG)\footnote{not to be confused with the relativistic MOND theory known as Tensor-Vector-Scalar gravity (TeVeS).}, is a gravitational paradigm which aims to describe the universe without dark matter. Although similar in goal, the general principles of MOG and MOND are entirely different. In MOG, the curvature of space-time is described by the metric tensor, $g_{\mu \nu}$ with the amount of curvature being affected by matter, a vector field, $\phi_{\mu}$, and two scalar field, $G$ and $\mu$, which represent a dynamical version of the Newtonian gravitational constant and the mass of the vector field respectively.

The detailed mathematics of MOG will not be discussed in this work, but rather just the significant result for spherical systems.  The modified gravitational law in spherical symmetry in MOG can be simply written as\citep{STVG}

\begin{equation}\label{MOGEqn2}
\nabla \Phi_{MOG} =\left(1+\alpha - \alpha\left(1+\mu r\right)\exp\left({-\mu r}\right)\right)  \frac{G M_{N}}{r^{2}}
\end{equation}

\noindent where $\alpha$ and $\mu$ are constants, $\alpha=8.89$ and $\mu = 0.042 \rm ~ kpc^{-1}$ and $M_{N}$ is the baryoninc mass. For radii $r\gg \mu^{-1}$, the approximate MOG acceleration law is

\begin{equation}\label{MOGEqn}
\nabla \Phi_{MOG} =\left(1+\alpha\right) \frac{G M_{N}}{r^{2}} = {\rm Constant} \times\nabla\Phi_{N}.
\end{equation}

It is clear that the MOG Eqn, \ref{MOGEqn} has the same form as the cluster regime of Eqn \ref{Justin_MOND_Cases}. Therefore, if the values of $A_{0}$ and $f$ are chosen correctly, the modified MOND paradigm  should be able to have the same profile as MOG for galaxy clusters.

\section{Emergent Gravity (EG)}\label{EGSec}

For completeness, we should show the result for EG gravity. The EG model works on the principle that gravity does not arise in the same manner as the other forces of nature (weak, strong and electromagnetic), but rather emerges as a result of entropy displacement \citep{EG}.

For a single body in spherical symmetry, the additional gravitational component predicted by EG can be written as

\begin{equation}\label{EG}
\nabla \Phi_{EG}(r) = \sqrt{\nabla\Phi_{N} a_{v}(r)}
\end{equation}

\noindent where $a_{v}(r)$ is determined from

\begin{equation}\label{EG2}
a_{v}(r) = \frac{a_{0}}{M_{N}(r)} \frac{d (M_{N}(r) r)}{dr},
\end{equation}

\noindent where in this equation $a_{0}$ is equivalent to the MOND acceleration scale\footnote{In previous EG work $a_{0}$ can be defined as $a_{0~EG} = c H_{0}$. In this case an extra factor of 1/6 appears in the equations as $a_{0} \approx c H_{0}/6$. The formulae we quote in Eqns \ref{EG} and \ref{EG2} is slightly simplified and is more in-line with our discussion on MOND.}. The hope is that the factor $a_{v}$ is high enough in galaxy clusters such that the dynamics can be explained without the need for particle dark matter. The next stage would be to understand how this law works beyond of spherical situations.

\section{Summary of Equations}\label{SummarySec}

When we perform our analysis of cluster A1689, we will be determining the success of the GMOND, EMOND, MOG and EG gravitational paradigms. We summarise the important formulae here. The total gravitational acceleration felt by a test particle for each of these paradigms are

\begin{equation}\label{summary}
\nabla\Phi = \begin{cases}

  \nabla\Phi_{N} + f \nabla \Phi_N & \text{GMOND clusters } \\

  \frac{1}{2}\left( \nabla\Phi_{N}+ \sqrt{\nabla\Phi_{N}}\sqrt{4A_{0}
(\Phi) + \nabla\Phi_{N}} \right) & \text{EMOND}\\
  (1+\alpha) \nabla\Phi_{N}  & \text{MOG clusters}\\
  
  \nabla\Phi_{N} + \sqrt{\nabla\Phi_{N} a_{v}(r)}& \text{EG}\\
  \nabla\Phi_{N} +  \nabla\Phi_{NFW} & \text{Newtonian.}

\end{cases}
\end{equation}

\section{A1689 Baryonic Mass Profile}\label{A1689}

Firstly, in order to model the cluster, we need to parametrise the baryonic mass contribution. The cluster is composed of two main sources of baryonic mass: galaxies, which dominate the centre and intra-cluster gas, which dominates the outer regions. These were modelled in \cite{A1689Paper} as

\begin{equation}\label{Galaxy}
\rho_{gal}(r) = \frac{M_{cg} (R_{co} + R_{cg})}{2\pi^{2} (r^{2} + R_{co}^{2})(r^{2} + R_{cg}^{2})}
\end{equation}

\noindent for the galaxies and, for the gas,

\begin{equation}\label{Gas}
n_{e~gas}(r) = n_{e0} \exp \left( k_{g} - k_{g} \left( 1 + \frac{r^{2}}{R_{g}^{2}} \right)^{1/(2 n_{g})} \right)
\end{equation}

\noindent where $\rho_{gas}(r) \approx 1.167 m_{p} n_{e~gas}$  such that  baryonic density is $\rho_{b}  =\rho_{gal} + \rho_{gas}$ and the total baryonic mass is $M_{b}(r) = \int^{r}_{0} 4\pi r'^{2} \rho_{b}(r') dr'$. In these equations, $M_{cg}$ is the mass of the galaxies, $R_{co}$ is the core radius of the galaxies, $R_{cg}$ is the radial extent of the galaxies, $n_{e0}$ is the central electron number density, $R_{g}$ is the radial extent of the gas and $n_{g}$ and $k_{g}$ are parameters which control the shape of the gas profile. The parameters which were used in \citep{A1689MOG}, which were taken from \cite{A1689Paper} are given in Table \ref{baryonparams}.

As the GMOND and MOG equations for total gravity essentially break down to a constant multiplied by the Newtonian gravity, to be successful it would be desirable for the Newtonian profile to be a similar shape to the NFW profile, defined above. We have little control over the shape of the gas profile as this quantity is determined from X-ray data. However, we do have some freedom to play with the galaxy profile.

We note that NFW profiles have a constant acceleration profile in the centre, meaning a $1/r$ density law or a mass growth proportional to $r^{2}$. From Eqn \ref{Galaxy}, the density profile is cored in the centre, which implies an associated Newtonian acceleration proportional to $r$. This is undesirable in MOG and GMOND as the total gravity is predicted to be a constant boost of the total Newtonian gravity.

\begin{figure}
\begin{tabular}{c}
\includegraphics[scale=0.7]{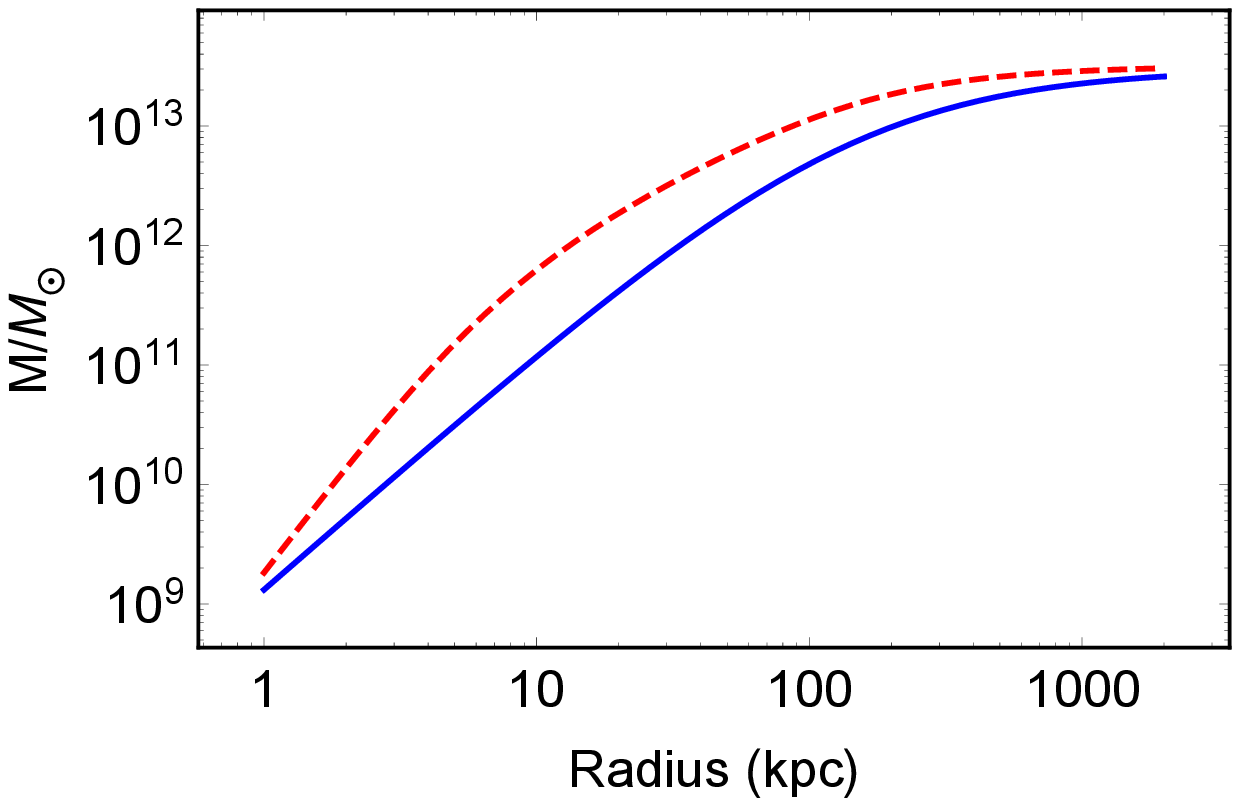}\\
\includegraphics[scale=0.7]{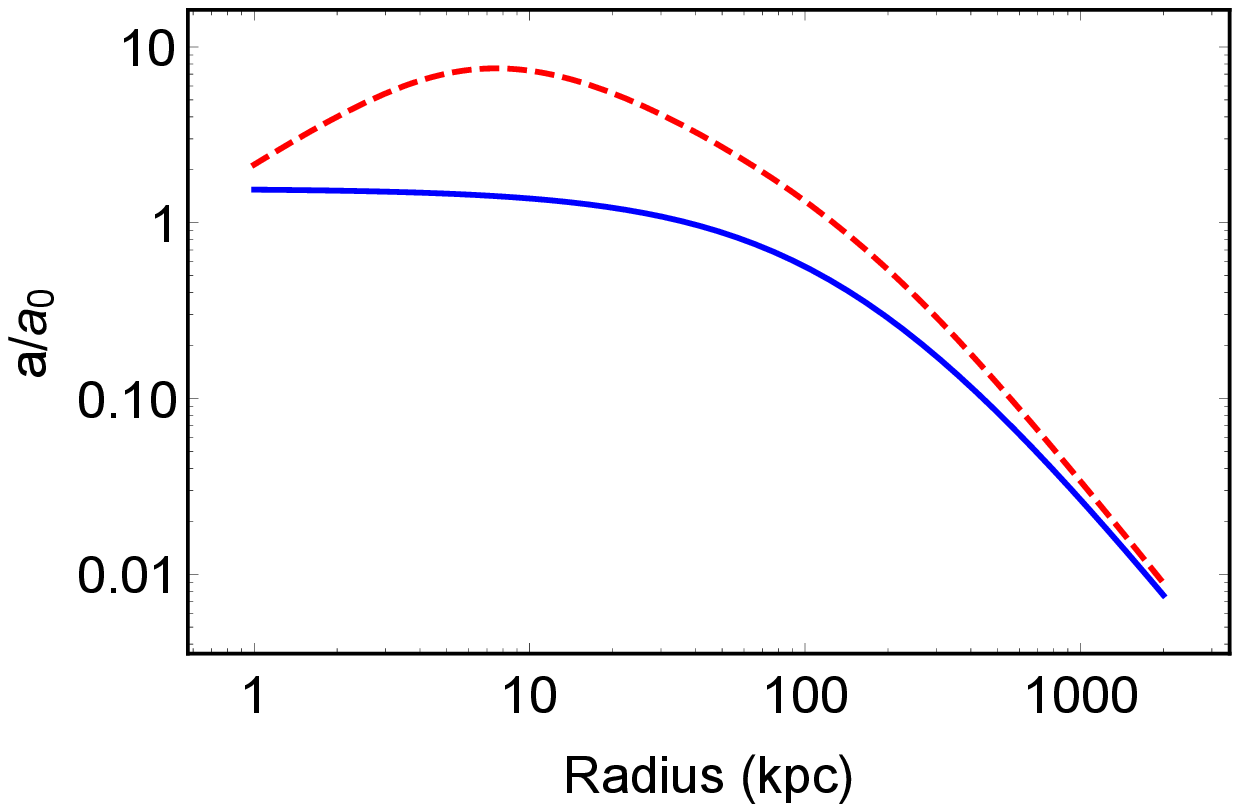}
\end{tabular}
\caption{Plot showing the mass (top panel) and acceleration (bottom panel) profile for a Hernquist galaxy model (blue) and the profile of Eqn \ref{Galaxy} (red, dashed). Parameters are as in Table \ref{baryonparams}. }
\label{GalMassPlot1}
\end{figure}

Interestingly, the Hernquist profile, which is described as \citep{hernquistprofile}

\begin{equation}
M_{H}(r) = \frac{M_{h} r^{2}}{\left( r + h \right)^{2}},
\end{equation}

\noindent where $M_{h}$ and $h$ are the total mass and radial extent of the galaxies respectively, exhibits the desired behaviour for the baryons in the centre of the cluster. It therefore only seems fitting to determine whether adjusting the galaxies mass profile can produce a better fit to A1689\footnote{In standard $\Lambda$CDM dynamics, the cluster dark matter profile dominates the mass, any effects of choosing different galaxy mass profiles would have little effect to the overall result. When trying to model systems without dark matter, the exact shape of the baryonic mass is indeed very important.}.  We have to determine what Hernquist parameters best match the general features of the NFW models from the literature. This is done empirically. We find that choosing a total baryonic mass of $M_{h} = 3 \times 10^{13} M_{\odot}$ and a Hernquist scale length of 150 kpc seems reasonable. We show, for comparison, the original baryonic mass profile as used in \cite{A1689Paper} and \cite{A1689MOG} and the new profile using the Hernquist galaxy model (Fig \ref{GalMassPlot1} top panel). We also show the acceleration profile (bottom panel).

\begin{table}[]
\centering
\caption{Table of parameters for the galaxy mass profile (Eqn \ref{Galaxy}) and the gas profile (Eqn \ref{Gas}) as taken from \citet{A1689MOG} and \citet{A1689Paper}.}
\label{baryonparams}
\begin{tabular}{|l|l|l|}
\hline 
Parameter & Value & Unit \\
\hline
$M_{cg}$ & $3.2\times 10^{13}$ & $M_{\odot} $             \\
$R_{co}$ & $5$                         & kpc                       \\
$R_{cg}$ & $150$                       & kpc                       \\
    &                           &                           \\
$n_{e0}$ & $0.0673$                    & cm$^{-3}$ \\
$R_{g}$  & $21.2$                      & kpc                       \\
$n_{g}$  & $2.91$                      & n/a                       \\
$k_{g}$  & $1.6$                       & n/a  \\
\hline                    
\end{tabular}
\end{table}

To be rigorous, we also test against a second gas profile from the literature as used by \cite[][]{sereno1,sereno2,A1689Um}. In these works, A1689 was modelled as a triaxial system, although spherical models were also computed. As non-spherical versions of EMOND, GMOND, MOG and EG have not been developed as of yet, we are forced to use the spherical results as an approximation. We propose that A1689 be re-analysed in modified gravity, dropping the assumption of sphericity, but this is beyond the scope of this work. 

\begin{figure}
\includegraphics[scale=0.7]{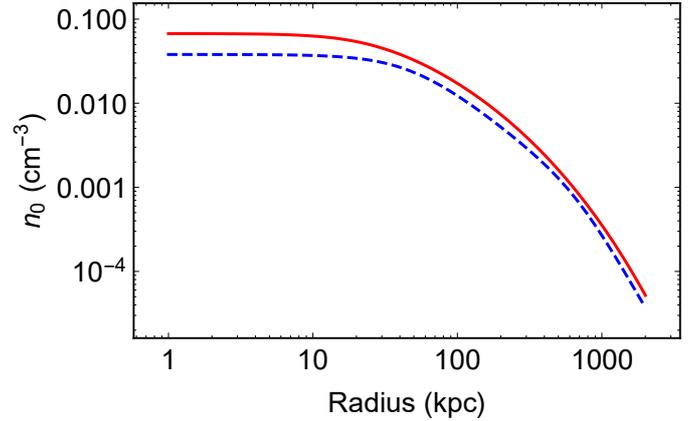}
\caption{Plot showing emission profiles following Eqn \ref{Gas} (Red) and Eqn \ref{gas2} (blue, dashed). Beyond $\approx 100$ kpc the profiles are similar. Within $\approx 100$ kpc the galaxies dominate the gas so we expect any differences associated with the choice of gas profile to be small. }
\label{Emission}
\end{figure}

The emission profile used in \citet[][]{sereno1,sereno2,A1689Um} follows 

\begin{equation}\label{gas2}
n_{e}(r) = n_{0} \left( 1 + \left( \frac{r}{r_{c}} \right)^{2} \right)^{-3\beta/2} \left( 1 + \left(  \frac{r}{r_{t}}\right)^{3} \right)^{-\gamma/3}
\end{equation} 

\noindent with the mass density calculated in the same way as previously. In Eqn \ref{gas2}, $r_{c}$ and $r_{t}$ are scale lengths and $\beta$ and $\gamma$ are dimensionless parameters which control the slope of the profile\footnote{Numerical values for the parameters were obtained by correspondence with the author of \citet[][]{sereno1}.}. We compare the emission profiles of Eqns \ref{Gas} and \ref{gas2} in Fig \ref{Emission}.

We can see the profiles are very similar and therefore expect similar results by using either. The largest difference is in the central regions, but as the galaxies of the cluster dominate there, we do not expect to see any visible difference in the result within $\sim 100$ kpc. In the outer part of the cluster, we expect the predicted total gravity by the modified theories to be slightly weaker when using the profile predicted by using Eqn \ref{Gas}. 

An additional test of this would be to model the cluster as a triaxial system.
\section{Results}\label{Results}

\subsection{EMOND boundary potential}

As the EMOND paradigm is dependant on the potential depth of the cluster, we have to rely on selecting an appropriate boundary condition, $\Phi(r_{200}) = \Phi_{\rm bound}$, to solve Eqn \ref{EMONDPoiss2}. Ideally, this boundary condition would be fixed by cosmological values very far away from a given source. Due to limitations of the simple model which we use, there are many factors which mean we cannot know what this factor is for each individual cluster\footnote{One of the key problems is that gas profiles in galaxy clusters tend to be described by divergent profiles, i.e. as $r\rightarrow \infty$ the enclosed gas mass does not tend to a constant. This makes analysis far away from the cluster difficult. } and therefore must estimate it empirically. A future study would have to be conducted to determine if the values for the boundary potential preferred by the EMOND paradigm are compatible with full scale simulations. This might be possible to do with the RAMSES code \citep{RAMSES} which has been given a MOND patch \citep{RAYMOND,PoR}. 

\subsection{GMOND Parameter Choice}

Before we can model A1689 in GMOND, the parameters $f$ and $A_{0}$ have to be fixed. If we work on the basis that we want to recreate the MOG profile in the GMOND paradigm, we have a constraint on $f$, $f = 1+\alpha \approx 10$. If we then assume that the value for $f$ is exactly this, we want to force the modified MOND equation into the cluster regime of Eqn \ref{Justin_MOND_Cases}. Therefore, we have a constraint on $A_{0}$, $A_{0} \gg f^{2}\nabla \Phi_{N}$. We can see from Fig \ref{GalMassPlot1} that the Newtonian acceleration profile for our Hernquist galaxy model in the central 50 kpc is approximately $1.5 a_{0}$. Therefore, our inequality can be rewritten as  $A_{0} \gg 1.5 f^{2} a_{0} $ or, using our $f \approx 10$ assumption, $A_{0} \gg 150 a_{0}$. This then gives an idea what values of $A_{0}$ we can choose to be in the cluster regime of Eqn \ref{Justin_MOND_Cases}. 

\subsection{Modified Gravity vs Newtonian}

\begin{figure}
\includegraphics[scale=0.7]{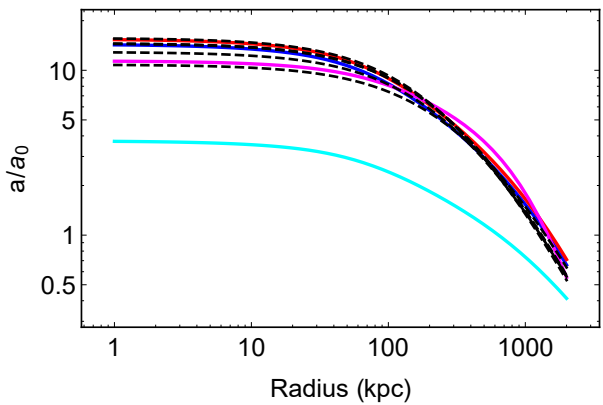}
\caption{Plot showing the predicted acceleration profiles for EMOND (magenta), GMOND (blue), MOG (red) and EG (cyan). This plot assumes A1689 to have a galaxy profile described by a Hernquist profile of mass $\rm 3\times 10^{13} M_{\odot}$ and scale radius 100 kpc and a gas profile described by Eqn \ref{Gas}. We compare to 5 NFW profiles from the literature (black dashed).}
\label{Acc1}
\end{figure}

Using Eqn \ref{summary}, we can compare the predictions of modified gravity with the best fit NFW profiles of A1689 from the literature. In Fig \ref{Acc1} we show the acceleration profile of A1689 for GMOND (blue), EMOND (magenta), MOG (red) and EG (cyan), comparing them to the five  NFW profiles (black dashed) described in Section \ref{NewtModel}.  For this, we choose the gas profile of Eqn \ref{Gas} and the Hernquist galaxy profile.

We can see that GMOND, EMOND and MOG all seem to be consistent with the NFW profiles. However, as mentioned in \cite{A1689Paper}, the EG prediction is too small. For this plot, we chose, for GMOND, $A_{0} = 200a_{0}$, and for EMOND $\Phi(r_{200}) = 1.5 \times 10^{12} \rm m^{2}s^{-2}$.\footnote{For the COMA cluster analysed in HZ2017, a boundary potential of $1.5 \times 10^{12} \rm m^{2}s^{-2}$ fits the overall cluster better, but a slightly higher boundary potential gave a better result for the UDGs. The result was the EMOND mass profile was higher than NFW predicted in for radii greater than approximately 1 Mpc. This could in theory be consistent with error bars in this region or gas profiles not being accurate far away from the cluster centre.}

In Fig \ref{Acc1}, we can see that the EMOND line is the correct shape, but may be slightly low in the region < 100 kpc. This might be an indication that EMOND prefers a larger galaxy mass profile. In order to determine what mass is preferred, a full $\chi$-squared analysis would be needed in conjunction with strong and weak lensing data, not attempted here.

\begin{figure}
\includegraphics[scale=0.7]{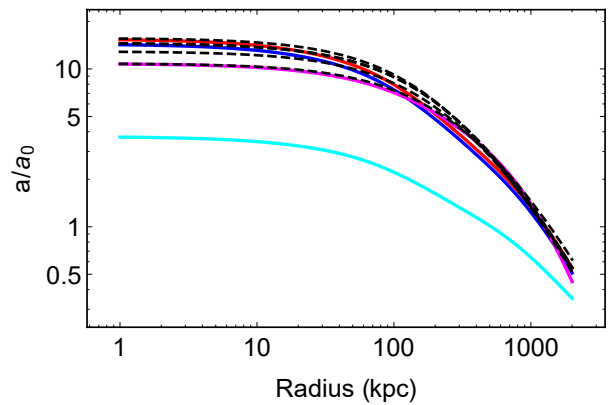}
\caption{Same as Fig \ref{Acc1} except using the gas profile of Eqn \ref{gas2} rather than Eqn \ref{Gas}.}
\label{Accgas2}
\end{figure}

As mentioned previously, two different gas profiles are used to model A1689. We show the effect of using the gas profile of Eqn \ref{gas2} instead of Eqn \ref{Gas} in Fig \ref{Accgas2}. In this case we see, as expected a small decrease in the acceleration profile in the outer regions with no discernible difference in the inner regions. The slight differences are almost negligible. 

\begin{figure}
\includegraphics[scale=0.7]{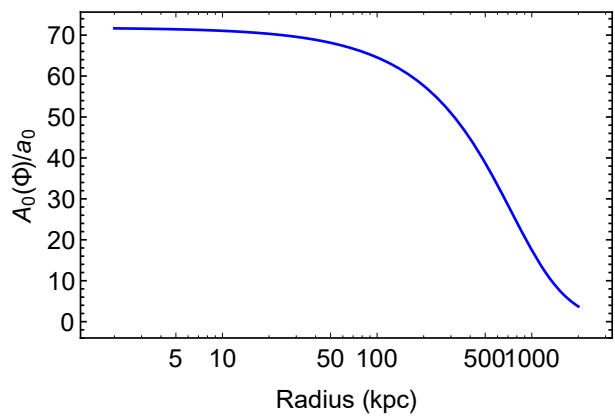}
\caption{Plot showing the value of $A_{0}(\Phi)$ as a function of radius in the EMOND paradigm. The EMOND parameter choices are as described in the text. This particular case is for the gas model of Eqn \ref{gas2}, but the gas model of \ref{Gas} does not change the result much. }
\label{A0Prof}
\end{figure}

For clarity, we also show the $A_{0}(\Phi)$ profile for the EMOND paradigm (Fig \ref{A0Prof}). The predicted gravitational potential in A1689 by EMOND is sufficiently large such that $A_{0}(\Phi) \approx 70 a_{0}$ in the central regions, allowing for a large boost of gravity compared to regular MOND. We should also stress that that the parameters used for this cluster are the same as was used in HZ2017 for Coma which shows consistency for the paradigm.

We next analyse why EG did not succeed in this cluster. Equation \ref{EG2} is fully analytic for a Hernquist model such that $a_{v}(r) =  a_{0}\left(1 + \frac{2 h}{r+h}\right)$. Therefore in the central regions of A1689, where the galaxies dominate the gas, the effective $a_{0}$ in EG can only be boosted by at most a factor of 3 This means that the Eqn \ref{EG} can at most be equivalent to $\nabla\Phi_{EG} = \sqrt{3}\sqrt{\nabla\Phi_{N} a_{0}}$. As the Newtonian acceleration is approximately  equal to $1.5 a_{0}$ in the centre, $\nabla\Phi_{EG} \approx 2 a_{0}$ and thus the total acceleration in EG ($\nabla\Phi_{N} + \nabla\Phi_{EG}$) is $ \approx 3.5 a_{0}$ in this region, which is what Fig \ref{Acc1} shows. To fit the A1689 cluster EG would need a boost of around 10, requiring a galaxy mass growth a lot steeper than the Hernquist profile, which seems unphysical.

\begin{figure}
\includegraphics[scale=0.7]{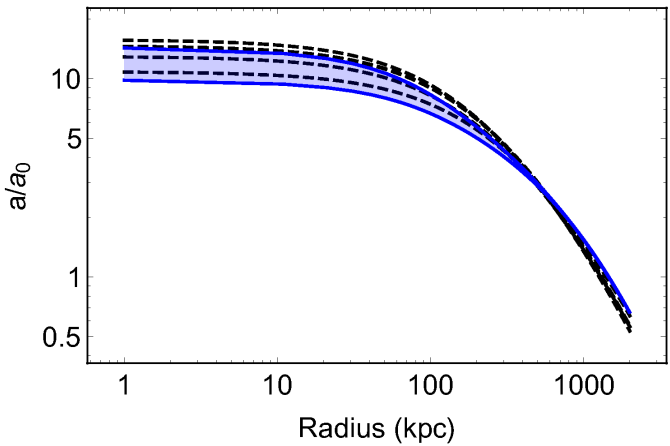}
\caption{Plot of the acceleration profile for GMOND showing how changing the value of $A_{0}$ from $50 a_{0}$ to $200 a_{0}$ changes predicted acceleration profile (blue shaded region). Lowest bound refers to lowest value of $A_{0}$. Again, we compare to the NFW profiled (black dashed lines). }
\label{Acc3}
\end{figure}

The final piece of analysis is to determine the role of $A_{0}$ in GMOND. Currently we have forced GMOND to reproduce a MOG-like curve by selecting $A_{0}$ such that we are in the cluster regime of Eqn \ref{Justin_MOND_Cases}. We should determine what happens if we select a smaller value of $A_{0}$ forcing GMOND into an intermediate regime. In Fig \ref{Acc3} we show a GMOND acceleration band for $A_{0}$ between $50a_{0}$ and $200a_{0}$ and over-plot the NFW profile for comparison.

We can see from Fig \ref{Acc3} that adjusting $A_{0}$ effectively acts like the concentration parameter in $\Lambda$CDM. The GMOND profile is slightly different from the NFW profile shape. An interesting study would be to determine the ideal parameter choice for A1689 looking at weak and strong lensing constraints both separately and combined. This is beyond the scope of this paper.   

\section{Conclusion}\label{Conclusion}

In this work, we modelled the galaxy cluster A1689 in different theories of modified gravity and compared the result to the best fit NFW profiles from the literature. Specifically we looked at EMOND, MOG, EG and what we call GMOND. Our results show that GMOND, EMOND and MOG all produce reasonable acceleration profiles, resembling the NFW profiles from the literature, but EG does not.

One point of note which should be stressed is the importance of the baryonic mass distribution. Unlike $\Lambda$CDM, where the overall mass is determined almost entirely by the dark matter, in modified theories of gravity the exact form of the baryons makes a significant difference. In the case of GMOND and MOG, due to the nature of the equations, the total acceleration is essentially a constant times the Newtonian potential. A similar argument can be made for EG as well. Therefore, both the shape and magnitude can affect the results significantly. This is a key consideration when modelling any astrophysical system in modified gravity, as we have shown here.

The next stage would be to perform a full and rigorous lensing analysis of A1689 in the context of EMOND, GMOND and MOG to determine whether they are consistent with the data. This would require much more work and is left for a future study. As well as invoking a lensing test, another avenue of research which could improve the modelling would be moving away from spherical symmetry. Currently, EMOND, GMOND and MOG have yet to be tested outside of spherical symmetry. Given the proposed triaxial nature of A1689 \citep{A1689Um}, breaking the assumption of sphericity is a natural step.

So far we have a consistency between the A1689 boundary potential and that used in the Coma cluster study of HZ2017. To make the EMOND conclusions more concrete, an understanding of the boundary potential has to be formulated. This would require an understanding of cosmology and examining the linear regime of the equations. 

For GMOND, we have not demonstrated any physical mechanism for choosing the $f$ and $A_{0}$ parameters, we have just empirically determined what values seem to give realistic results for the cluster A1689. The obvious next stage is to try fit a large sample of galaxy clusters to determine  the best fit $f$ and $A_{0}$ parameters with the intention of finding a dynamical mechanism which can naturally explain how they change between galaxies and galaxy clusters. Clearly in galaxies, $A_{0}$ should be equivalent to $a_{0}$ and $f$ should be $\gtrapprox 10$. Galaxy cluster constraints will require numerical testing. One possible route for making at least $A_{0}$ dynamical would be to combine this paradigm with EMOND. This would make the equations a little more complicated. We are unsure what mechanism would best work for making $f$ dynamical at this stage.

What we demonstrate in this paper (which has also been demonstrated in \cite{A1689Paper}) is that although A1689 is an issue in current and more mainstream theories of modified gravity, there are some avenues of research which need to be explored before we rule out modified gravity without dark matter on the cluster scale. We do acknowledge that the solution may not be as mathematically elegant as Newtonian physics with dark matter but would links between the total observed gravity and the contribution of visible matter.

\section*{Acknowledgements}

We would like to thank Theo Nieuwenhuizen and John Moffat for comments on the draft and Mauro Sereno for providing discussions on the X-ray gas contribution.  AOH is supported by Science and Technologies Funding Council (STFC) studentship (Grant code: 1-APAA-STFC12).

%%%%%%%%%%%%%%%%%%%%%%%%%%%%%%%%%%%%%%%%%%%%%%%%%%

%%%%%%%%%%%%%%%%%%%% REFERENCES %%%%%%%%%%%%%%%%%%

% The best way to enter references is to use BibTeX:

\bibliographystyle{mnras}
\bibliography{A1689bib} % if your bibtex file is called example.bib

%%%%%%%%%%%%%%%%%%%%%%%%%%%%%%%%%%%%%%%%%%%%%%%%%%

%%%%%%%%%%%%%%%%% APPENDICES %%%%%%%%%%%%%%%%%%%%%

\appendix

\section{A toy model of GMOND galaxy dynamics}

\begin{figure}
\begin{tabular}{c}
\includegraphics[scale=0.6]{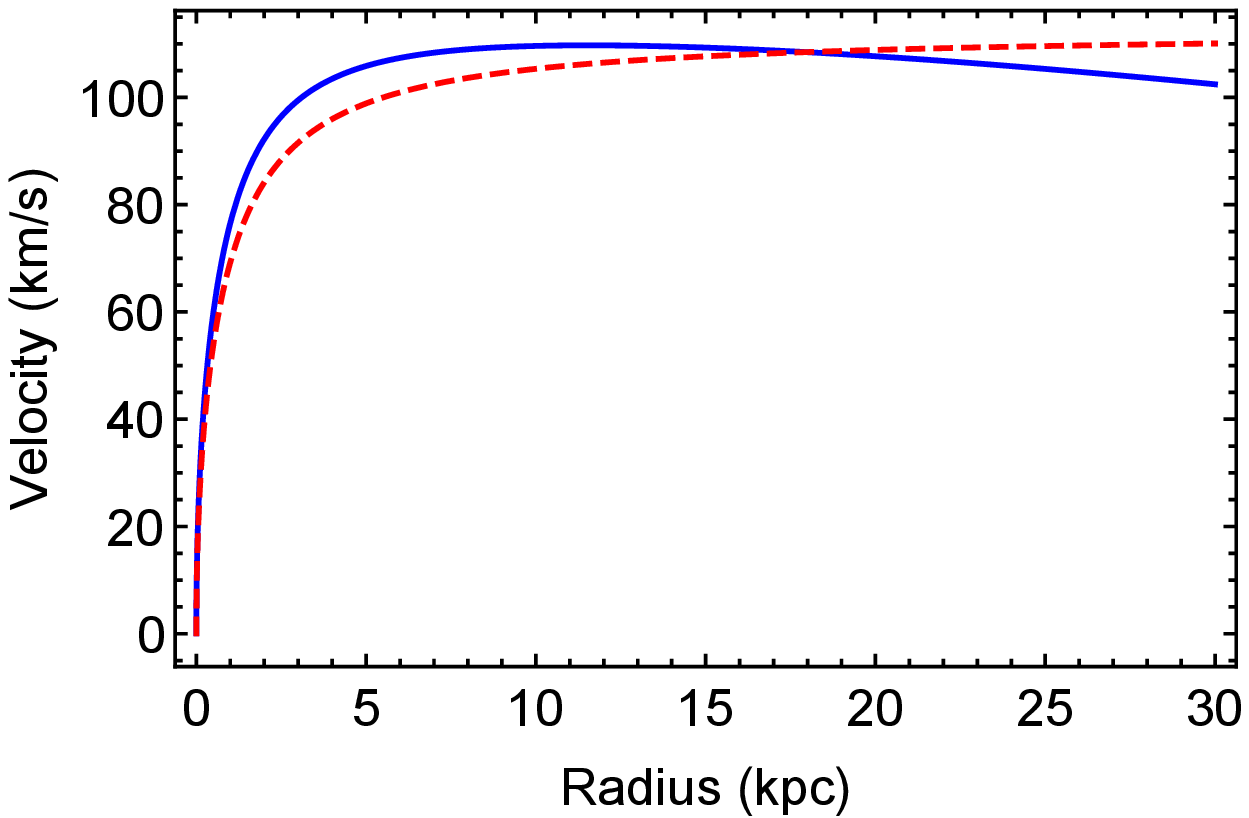}\\\includegraphics[scale=0.6]{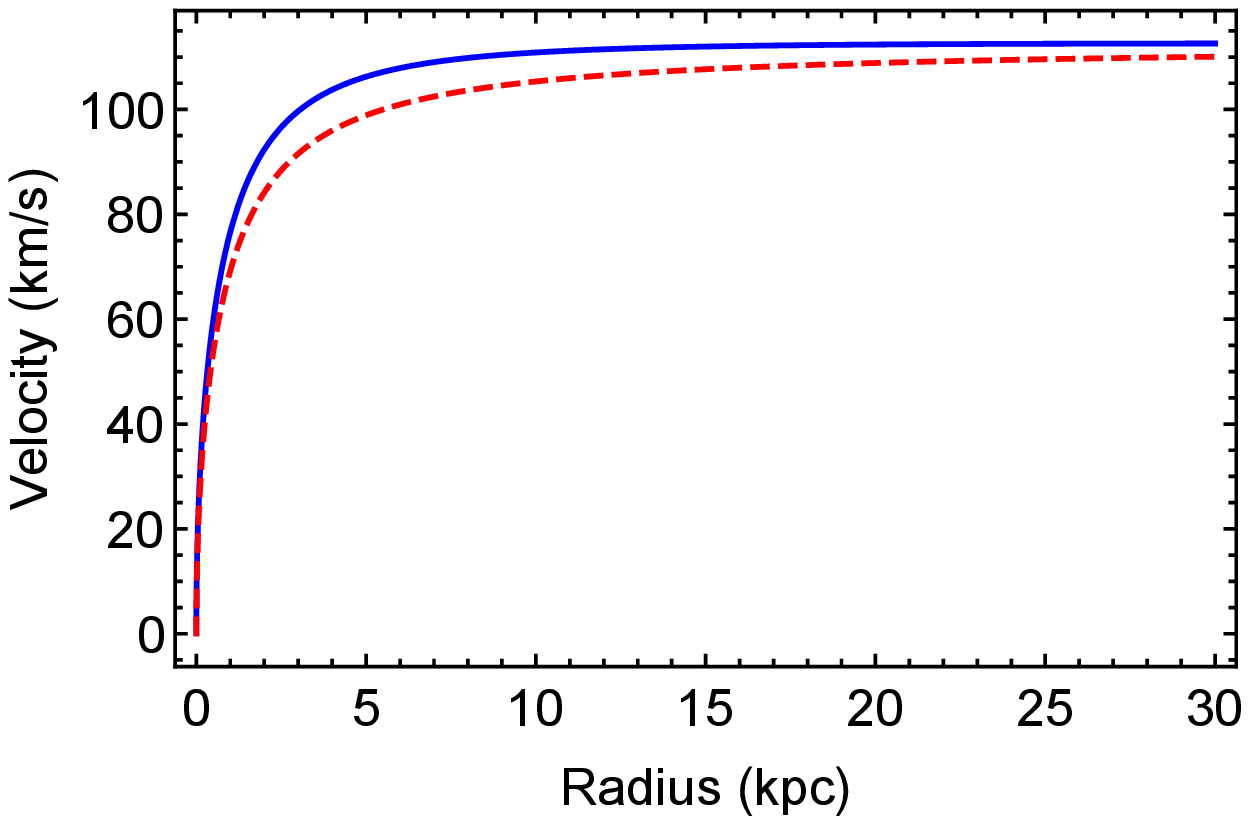}
\end{tabular}
\caption{Plot showing the predicted rotation curve in GMOND (blue curve) and regular MOND (red dashed line).  Top panel has GMOND parameters $A_{0}=a_{0}$ and $f = 10$. Bottom panel has GMOND parameters $A_{0}=a_{0}$ and $f = 100$. Both curves have baryonic mass $M_{h} = 10^{10}~ \rm M_{\odot}$ and scale length $h = 3 ~ \rm kpc$.}
\label{RotCurve}
\end{figure}

For an illustration of the GMOND paradigm, we make a toy model galaxy and show the predicted rotation curve from Equation \ref{Justin_Poisson}. To model the toy galaxy, we choose a Hernquist matter distribution, which follows \citep{hernquistprofile}

\begin{equation}
M_{H}(r) = \frac{M_{h} r^{2}}{\left( r + h \right)^{2}} 
\end{equation}

\noindent where $M_{h}$ is the Newtonian mass of the system and $h$ is the galaxy scale length. The rotation curve in the Modified MOND paradigm is defined as $v_{r}(r) = \sqrt{r~\nabla\Phi}$ with $\nabla\Phi$ defined as the gradient of Eqn \ref{Total_Potential}. We show the rotation curves for this paradigm (blue curve) and regular MOND with a simple interpolation function, ($\mu(x) = 1/(1+x)$) \citep{simplemu}, (red dashed curve) in Fig \ref{RotCurve}.

\begin{figure}
\includegraphics[scale=0.7]{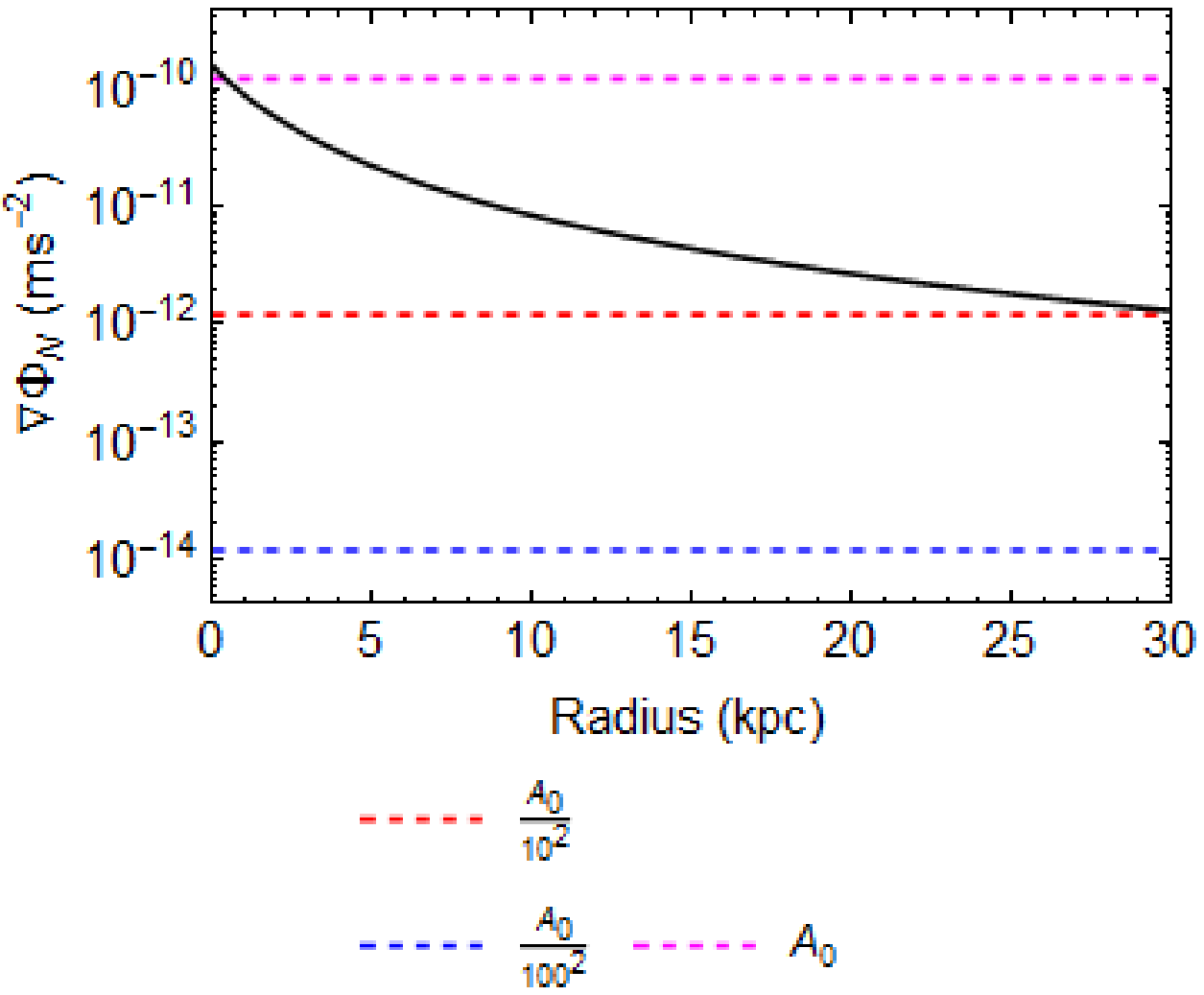}
\caption{Plot showing the Newtonian gravitational acceleration as a function of radius. We compare the value to our chosen $A_{0} = a_{0}$ and the value of $A_{0}/f^{2}$ for $f=10$ (red dashed line) and $f=100$ (blue dashed line). This plot demonstrated the transitioning between GMOND regimes.}
\label{RotCurvedemo}
\end{figure}

In this plot, we show rotation curves for two sets of parameter choices; the top panel has $f=10$ and the bottom panel has $f=100$. Both rotation curves have $A_{0}=a_{0}$. In the bottom panel, the Modified MOND curve is aesthetically similar to the regular MOND curve, but slightly higher. The reason for the same shape is that increasing $f$ forces Eqn \ref{Justin_MOND_Cases} to become closer to the Deep-MOND limit rather than an intermediate case. We demonstrate this in Fig~\ref{RotCurvedemo} where we show the Newtonian acceleration compared with the value of $A_{0}$ and $A_{0}/f^{2}$.

In this figure, we can see that between $\approx 5$ and $\approx 15$ kpc, the Newtonian acceleration is less than $A_{0}$ and greater than $A_{0}/f^{2}$, and the galaxy should therefore behave like  deep-MOND. Greater than $\approx 15$ kpc, the Newtonian acceleration is still much greater that $A_{0}/f^{2}$ for the $f=100$ case, but is comparable for the $f=10$ case. Therefore when $f=10$, the outside of the galaxy is in-between the deep-MOND and cluster regime of GMOND and hence the rotation curve falls with radius, but less so than a pure Newtonian case\footnote{Rotation curves which decline in the outer radii are in fact in line with some observed galaxies. Declining rotation curve galaxies have been tested in the MOND paradigm in \cite{declining_curves}. The falling rotation velocity is achieved by the external field effect (EFE). They find that, within the errors, most galaxies can be explained by the external field effect. An interesting constraint on the value of $f$ in GMOND might be galaxies of this kind. }.

The reason that the curve is slightly higher for the $f=100$ case in Fig \ref{RotCurve} than the regular MOND prediction is the total gravitational potential in GMOND is composed of the scalar field and the Newtonian potential, whereas in MOND, there is no additional Newtonian potential term. A sample of galaxies with rotation curve data could be easily fit in GMOND to try and constrain $f$. As MOND works well with galaxies, high $f$ values should be accepted, but a lower limit should be determined. This is beyond the scope of this paper.

\section{Analytical solution in spherical symmetry}\label{AppAnalytic}

We show the analytical solution for the scalar field in the GMOND paradigm assuming spherical symmetry. The scalar field Poisson Equation, Eqn \ref{Justin_Poisson}, can be integrated in spherical symmetry,

\begin{equation}\label{phiPoissonspherical}
F'(Y) \frac{\rm d \tilde{\phi}}{\rm d r} = \frac{GM(r)}{r^{2}}.
\end{equation}

\noindent which can then be analytically solved for $\tilde{\phi}'(r)$. Remembering the expression for $F(Y)$, given in Eqn \ref{Ffunc}, $F'(y)$ is

\begin{equation}
F'(Y) = \frac{1 - 3 \mathcal{T}_{1} Y}{2 f \sqrt{1 - \mathcal{T}_{1} Y}},
\end{equation} 

\noindent where $\mathcal{T}_{1} \equiv \left( 2f/3 \right)^{2}$. Using this, Eqn \ref{phiPoissonspherical} can be rearranged to,

\begin{equation}\label{phiPoissonspherical2}
\frac{\rm d \tilde{\phi}}{\rm d r} = \sqrt{-\frac{4 \mathcal{T}_{2}}{9\mathcal{T}_{1}} + \frac{2^{5/3}\mathcal{T}_{2}^{2}}{9\mathcal{T}_{4}^{1/3}} + \frac{2^{1/3}\mathcal{T}_{4}^{1/3}}{9\mathcal{T}_{1}^{2}} + \frac{2^{5/3} \mathcal{T}_{3}^{2}\mathcal{T}_{2} \mathcal{T}_{1} f^{2}}{3 \mathcal{T}_{4}^{1/3}}}
\end{equation}

\noindent where

\begin{align}
\mathcal{T}_{2} &=  a_{0}^{2}\\
\mathcal{T}_{3} &= \frac{GM(r) }{r^{2}}\\
\mathcal{T}_{4} &= 4\mathcal{T}_{2}^{3}\mathcal{T}_{1}^{3} + 3\sqrt{3} \mathcal{T}_{5} + 45 \mathcal{T}_{3}^{2} \mathcal{T}_{2}^{2} \mathcal{T}_{1}^{4} f^{2}\\
\mathcal{T}_{5} &= 8 \mathcal{T}_{3}^{2}\mathcal{T}_{2}^{5} \mathcal{T}_{1}^{7} f^{2} + 59 \mathcal{T}_{3}^{4} \mathcal{T}_{2}^{4} \mathcal{T}_{1}^{8} f^{4} - 16 \mathcal{T}_{3}^{6} \mathcal{T}_{2}^{3} \mathcal{T}_{1}^{9} f^{6}.
\end{align}

Equation \ref{phiPoissonspherical2} can be solved to determine the contribution of the scalar field to the gravitational potential. We note that this is a solution to a cubic equation and care should be taken in choosing the correct branch of the solution.

%%%%%%%%%%%%%%%%%%%%%%%%%%%%%%%%%%%%%%%%%%%%%%%%%%

% Don't change these lines
\bsp	% typesetting comment
\label{lastpage}
\end{document}